\documentstyle[12pt]{article}

\newcommand{\beq}{\begin{equation}}
\newcommand{\eeq}{\end{equation}}
\newcommand{\bea}{\begin{eqnarray}}
\newcommand{\eea}{\end{eqnarray}}
\newcommand{\ben}{\begin{enumerate}}
\newcommand{\een}{\end{enumerate}}
\newcommand{\bit}{\begin{itemize}}
\newcommand{\eit}{\end{itemize}}
\newcommand{\ba}{\begin{array}}
\newcommand{\ea}{\end{array}}

\newcommand{\mb}{\bf}

\newcommand{\al}{\alpha}

\newcommand{\lm}{\lambda}

\newcommand{\rw}{\rightarrow}

\newcommand{\de}{\partial}

\newcommand{\ti}{\tilde}
\newcommand{\x}{\mbox{X}}

\newcommand{\eq}{\begin{equation}}
\newcommand{\en}{\end{equation}}
\newcommand{\eqa}{\begin{eqnarray}}
\newcommand{\ena}{\end{eqnarray}}

\newcommand{\PRL}[1]{Phys.\ Rev.\ Lett.\ {\bf #1}}

\newcommand{\TMP}[1]{Teor.\ Mat.\ Fiz.\ {\bf #1}}

\begin{document}

\hskip 11.5cm \vbox{\hbox{DFTT 7/2003}\hbox{April 2003}}
\vskip 0.4cm

\vskip 1.5cm
\centerline{\Large\bf A new class of solutions of the DMPK equation.}
\vskip 1cm
\centerline{M.Capello$^a$ and M. Caselle$^b$}
\vskip .3cm
\centerline{\sl $^a$ SISSA, via Beirut 2-4, I-34014 Trieste, Italy} 
\centerline{$^b$  Dip. di Fisica 
Teorica dell'Universit\`a di Torino and I.N.F.N.}
\centerline{via P.Giuria 1, I-10125 Turin,Italy}

\vskip .5cm

\vskip 1.5cm

\begin{abstract}
We introduce and discuss a new class of solutions of the DMPK equation in which
some of the eigenvalues are grouped into clusters which are
conserved  in the asymptotic large distance limit (i.e. as the length of the
wire increases). We give an explicit expression for the asymptotic expansion of
these solutions and suggest some possible applications. In particular these new
solution could be useful to avoid the quasi one dimensional constraint
 in the DMPK equation and to study the
crossover between the metallic and insulating phases. 
\end{abstract}
\vskip 0.3cm
\noindent


\newpage

\setcounter{footnote}{0}
\def\thefootnote{\arabic{footnote}}

\section{Introduction}
One of the most interesting tools to describe the electron transport properties
of quantum wires is the so called Dorokhov Mello Pereyra Kumar (DMPK)
equation~\cite{dmpk}. This equation describes the evolution of the joint
probability distribution of the transmission eigenvalues $P(\{\lambda_i\})$
as the length of the
wire increases and has been the subject of intense study in these last
years~\cite{caselle1}-\cite{yyy}. Among the several remarkable features of this equation, the 
most interesting one is that it can be exactly mapped onto the radial part of
the Laplace-Beltrami operator of suitable symmetric spaces~\cite{caselle1}.
Thanks to this mapping it is possible to write the Green function of the DMPK
equation in terms of the so called Zonal Spherical Functions (ZSF) of the
corresponding symmetric spaces. In the
 $\beta=2$ 
case the ZSF can be written explicitly in terms of ordinary ipergeometric
functions~\cite{bk} thus leading to an exact 
solution of the DMPK equation~\cite{br} while for $\beta=1,4$ one can 
rely on a powerful asymptotic
expansion due to Harish-Chandra~\cite{hc}. This approach leads to a rather
involved  expression for $P(\{\lambda_i\})$ which however drastically simplifies in
the two limits of very short wires (metallic regime) and very long wires
(insulating regime), thus allowing to evaluate all the quantities of interest
(see~\cite{br} and~\cite{beenakker1} for a
detailed discussion). 

However, despite these remarkable results there
are two major drawbacks in the DMPK approach to quantum wires. The first is that
the DMPK description only holds in the (quasi) one dimensional limit. The second
is that the above mentioned solution (even in the simpler $\beta=2$ case)
 does not allow to study 
the intermediate region between the metallic and the insulating
regimes (the so called ``cross-over regime'')  where no simplifying 
approximation is allowed. This region has recently attracted much interest since
both numerical and analytical results seem to indicate a rather non-trivial
(with possibly a non-analytic point~\cite{mwgg}) 
behaviour of $P(\{\lambda_i\})$ exactly in this regime~\cite{zzz,gm,gmw}.

In these last years several interesting approach have been proposed to overcome
these problems. Various generalizations of the DMPK equation have been
suggested~\cite{yyy} \cite{mello} to avoid the quasi one dimensional limit. However in all
these generalized equations most of the nice properties of the DMPK equation are
lost (mainly due to the fact that the description in terms of
symmetric spaces is lost) and only few informations on the expected joint
probability density of the transmission eigenvalues can be obtained.
The aim of this paper is to propose a completely different approach to these
problems, which instead fully exploits the power of the symmetric space
structure which is behind the DMPK equation. Instead of modifying the DMPK
equation , we shall keep it unchanged, but shall look for a set of special
solutions (with non trivial initial conditions) which break the isotropy ansatz.
To this end we shall use the fact
that the DMPK equation can be mapped into the evolution operator of a class of
1D quantum integrable models known as Calogero-Sutherland (C-S) models (for a
review see~\cite{olshanetsky}). As a
consequence of their exact integrability it is possible to show that in these
models,
besides the well known symmetric solution, a wide class of non
trivial (but exact) solutions exist in which the particles  are
 grouped into clusters which survive in the asymptotic limit.
 The clusters of particles become, once the mapping with the DMPK equation is
 performed, clusters of eigenvalues. The exact integrability of the C-S model
 ensures that this asymmetric distribution of eigenvalues survives in the
 asymptotic limit and the remarkable properties of the underlying symmetric
 space allow to write explicitly such asymptotic expansion.  

This paper is organized as follows:
In sect 2 we shall recall some known results on 
quantum transport, focusing on
the DMPK equation in order to
fix notations.
In sect 3 we shall review  the main concepts associated to  C-S models and the 
mathematical tools applied in order to find their solutions. 
 In sect 4 we shall describe the new class of solutions of the
DMPK equation and  in sect 5 we shall give some hints on how these new
solutions could be used in order to address the open problems mentioned in the
introduction.

\section{Quantum transport in a wire and  DMPK equation}
 
A mesoscopic conductor can be modeled as a disordered
region located between two ideal leads connected to two electron
reservoirs; at very low temperatures, frequencies and voltages
the scattering phenomena inside the wire are supposed to be elastic,
and the electron wave functions are assumed to be phase coherent.
Given a quantum wire of lenght $L$ and width $W$ ($L\gg W$) it is
possible to find a finite but large number $N$ of Fermi channels
associated to the conduction electrons.  Adopting the notation used in \cite{dmpk}, we call $I$ and $O$ the $N$-dimensional vectors associated
respectively to the amplitudes of the incoming and outgoing waves on
the left of the wire and $I'$ and $O'$ the corrisponding vectors on the
right.

The scattering phenomena inside the wire can be described by means of
the $2N$-dimensional Transfer matrix $M$,  which
connects the amplitudes of each channel located on the left to the ones
on the right of the wire:
\beq \label{eq: M transf}
M\left( \begin{array}{c} I \\ O \end{array} \right) =
\left( \begin{array}{c} O' \\ I' \end{array} \right)
\eeq
The matrix $M$ can be decomposed in a radial and an angular part
according to the polar decomposition \cite{dmpk}:
\beq
\label{eq:Mpara}
M=\left( \begin{array}{cc} u_1 & 0 \\
                           0 & u_3 \end{array} \right)
\left( \begin{array}{cc} \sqrt{1+\Lambda} & \sqrt{\Lambda} \\
                         \sqrt{\Lambda} & \sqrt{1+\Lambda}\end{array}
 \right)
\left( \begin{array}{cc} u_2 & 0 \\
                         0 & u_4 \end{array} \right) \equiv U\Gamma V
\eeq

with $u_i$ unitary matrices $N\times N$, which assume the role of
angular coordinates and $\Lambda$ real diagonal matrix
with non negative elements $\{\lm_1\ldots \lm_N\}$, as
radial coordinates.

In \cite{dmpk}  the
general symmetries of the system (time reversal and/or  rotational
symmetry) fix  the matrix $M$ to a particular Lie group:
\bit
\item
if time reversal symmetry (TRS) is present and there is no
spin-rotational symmetry (SRS) inside the wire $M \in Sp\ (2N,\mb R)$;
\item if time reversal symmetry and spin-rotational
symmetry are
both present  inside the wire $M\in SO^*(4N)$;
\item if time reversal symmetry is absent $M\in SU(N,N)$.
\eit
Experimentally these three symmetry classes correspond to the presence or
absence of a magnetic field (which breaks the time reversal
symmetry) and to the relevance of the spin-orbit interaction term
(which breaks spin-rotational symmetry) inside the wire.

The general symmetries of the Transfer matrix lead to a possible
formulation of the problem by means of the tools offered by 
Random Matrix Theory; each symmetry class is
assigned to an ensemble of tranfer matrices, associated to a
parameter $\beta$ which can be interpreted as the number of degrees of
freedom of each matrix element.

In the framework of the transfer matrix approach to 
quantum wires the DMPK equation \cite{dmpk} describes the evolution of the
probability distribution $P_L$ of the transmission
eigenparameters $\{\lm_i\}$  as a function of the reduced lenght $s=L/l$
($L$ being the length of
the wire and $l$ the mean free path associated to the electron): 
 \bea \label{eq:DMPK}
\frac{\partial P_L}{\partial s}&=&\frac{2}{\gamma }\, \sum_{i=1}^N
\frac{\partial }{\partial \lambda_i }\lambda_i (1+\lambda_i)\, J_\beta
\frac{\partial }{\partial \lambda_i }J_\beta^{-1} P_L\equiv D\ P_L
\\ \nonumber\\
\mbox{with  } \ \ \gamma &\equiv &\beta N +2-\beta \nonumber
\eea
where $N$ is the number of Fermi channels
associated to the conduction electrons, 
$\beta=\{1,\ 2,\ 4 \}$ is the parameter associated to the
symmetry of the system, $J_\beta $ is the Jacobian from the space of
the whole matrix $M$ to the space of the eigenparameters $\{\lm_i\}$:
\[J_\beta (\{ \lambda_i \}
)=\prod_{i=1}^N\prod_{j=i+1}^N|\lm_i-\lm_j|^\beta\] and $D$ is the DMPK
operator.
Given the probability distribution $P_L$ the conductance of the wire can be obtained by means of the
Landauer formula \cite{landauer}.
It is important to stress that the construction of the 
DMPK equation requires an {\em isotropy ansatz} among the channels. 
Physically this implies that
the finite time required for the electron to be scattered in the
transversal direction is assumed infinitesimally small, and this assumption
fits only for a quasi-1D wire.

The Transfer matrix ensembles which are at the basis of the DMPK equation
are very interesting from a mathematical point 
of view since they can be mapped on suitable  symmetric
space of {\bf negative} curvature (see ~\cite{caselle1} and \cite{review} for a
detailed discussion). Similar representations in terms of symmetric spaces also
exist for other Random Matrix Theories, but they usually involve symmetric
spaces of zero or positive curvature (see~\cite{review}). The negative curvature
of the space is a peculiar feature of the transfer matrix ensembles and has
relevant consequences on the quantum wires applications of these ensembles. For
instance it can be shown that
the eigenparameters $\{\lm_i\}$ appearing in formula
(\ref{eq:Mpara}) inside $\Lambda$ exactly correspond to the radial coordinates
 of the symmetric
space while the ``radius'' of the space can be mapped (with a suitable
normalization) onto the length of the wire. 
The fact that the radial coordinates in a negative curvature symmetric space
flow to infinity as the radius increases then becomes in the quantum wire
context the well known result that as the length of the wire increases the wire
moves from a conducting to an insulating regime. We shall discuss this mapping
in great detail in the next section.

Table~\ref{tab1} contains the main symmetry classes associated to
the parameter $\beta$ and the corresponding symmetric space $\x$ associated to the eigenparameters $\{\lm_i\}$.

\begin{center}
\begin{table}[!h]
\caption{
Symmetry classes for a quantum wire \label{tab1}}
\vskip5mm

\hskip .2cm
\begin{tabular}{|c|c|c|c|c|}
\hline
  $\beta $  & TRS
&SRS  & $M$ & $\x$ \\
\hline
& & & & \\
1 & Yes & Yes & $Sp(2N,\mb R)$ & $Sp(2N,\mb R)/U(N)$ \\
2 & No  & Yes & $SU(N,N)$      & $SU(N,N)/SU(N)\otimes SU(N) \otimes
U(1)$\\
4 & Yes & No  & $SO^*(4N)$     & $SO^*(4N)/U(2N)$\\
 & & & & \\
\hline
\end{tabular}
\end{table}
\end{center}

\section{Calogero-Sutherland models:
mapping of the DMPK equation on a symmetric space and solutions
\label{sec:calogero}}

In order to address the symmetric space description of quantum wires we need an
intermediate ingredient, i.e. the well known Calogero-Sutherland models (for a
review see~\cite{olshanetsky}). 
These models describe $N$ interacting particles in
one dimension, and are characterized by an Hamiltonian that can be mapped, for
certain values of the coupling constants,  into the radial part
of the Laplace-Beltrami operator ($B$ in the following)
on a suitable symmetric space $\x$.
This operator is the generalization of the familiar Laplace operator describing
 free diffusion in the  Cartesian space and  it  describes free diffusion 
on a generic manifold $\x$ (in our case $\x$ is one of the
 non-compact symmetric spaces $\x$ of Table \ref{tab1}, characterized by a  root lattice structure of type $C_N$).

The action of the Laplace-Beltrami operator can be formulated in terms of the following eigenvalue equation:
\beq B\Phi_k(x)=k^2 \Phi_k(x)\eeq
the eigenfunctions $\Phi_k(x)$ are known in the literature as ''zonal spherical
 functions'' (ZSF) (for a review see \cite{review}).

The connection between  the C-S Hamiltonian and the symmetric space (in particular the root
lattice associated to the symmetric space) and the mapping of the C-S Hamiltonian into the radial part
of the Laplace-Beltrami operator is the key point which determines 
the integrability of the model.

The general form of the Calogero-Sutherland Hamiltonian is:
\bea
\label{eq:calH}
{\cal H}=\frac{1}{2}\sum_{i=1}^n p_i^2 + \sum_{\alpha \in \Delta^+}
g_\alpha^2\, v(x_\alpha)\nonumber \\
x=(x_1,...,x_N),\ \ \ \ p_i=-i\frac{\partial }{\partial x_i},\ \ \ \
x_\alpha = (x, \alpha) \\ \nonumber
\eea

where  $\Delta^+$
is the set of positive roots  $\al$  associated to the space $\x$,
$(x,\alpha)$ denotes the scalar product,
 $v(x_\alpha)$ is the interaction potential
and $g_\al$ is a coupling constant which  depends on the roots as:
\beq
\label{eq:rootvalues}
g_\alpha^2=\frac{m_\alpha (m_\alpha -2)|\alpha^2|}{8}
\eeq
where $m_\alpha$ is the root multiplicity in the space $\x$ and
$|\alpha^2|$ its length; relation (\ref{eq:rootvalues}) is a necessary
 constraint for the model to be solvable. The particular C-S Hamiltonian 
 we are interested in contains a  potential of the form:
\beq v(x_\alpha)=\frac{1}{\sinh^2 x_\al}
\label{rev1}\eeq

The relevant point for the present analysis is that 
if we choose, as  in \cite{br}:
\beq \lm_i=\sinh^2 x_i\eeq
the DMPK operator can be mapped exactly onto a C-S Hamiltonian with a
potential of the type (\ref{rev1}) and the symmetries of one of the 
 symmetric spaces listed in Table~1.  

These spaces are characterized by root lattices (denoted as $C_N$)
 composed both by ordinary roots
(whose multiplicity $m_o=1,2$ or $4$ can be identified with the $\beta$ parameter which encodes
the symmetry properties of the matrix model) and long roots with multiplicity
$m_l=1$ which are responsible for 
the peculiar properties of the eigenvalues near
the boundary $\lambda\sim 0$ (see \cite{caselle1,review} for further
details). 
Given the set $\{e_j\}$ of linearly independent versors associated to the space,
ordinary roots are commonly written as $\al=\pm e_j \pm e_i$ ($j \neq i=1,N)$ and
long roots correspond to $\al=\pm 2e_j$ $(j=1,N)$.

These root lattices are characterized by a $Z_2$ symmetry, i.e. they
are invariant under reflection. This means that the origin plays a distinguished
role. Indeed, most of the results  we shall discuss in the following can be more
easily understood by introducing an additional (fictitious)
eigenvalue which is kept fixed in the origin and interacts (with a standard
repulsive interaction mediated by the long roots of the lattice) with all the 
other eigenvalues.

\begin{center}
\begin{table}[!h]
\caption{
Symmetric spaces associated to the transfer matrix and their root
multiplicities \label{tab2}} \vskip5mm

\hskip .2cm
\begin{tabular}{|c|c|c|c|c|}
\hline
  $\beta $  &
 $\x$ & $m_o$ & $m_l$ & $m_s$ \\
\hline
& & & & \\
1 & $Sp(2N,\mb R)/U(N)$ & 1 & 1& 0\\
2 & $SU(N,N)/SU(N)\otimes SU(N) \otimes
U(1)$ & 2 & 1& 0\\
4 & $SO^*(4N)/U(2N)$ & 4 & 1 & 0\\
 & & & & \\
\hline
\end{tabular}
\end{table}
\end{center}

The complete serie of steps involved in this
mapping can be found in reference~\cite{caselle1,review}.
In particular it turns out that the radial part of the
Laplace--Beltrami operator $B$ on the symmetric space $\x$ is related
to the DMPK evolution operator $D$ in equation~(\ref{eq:DMPK}) by:
\beq
\label{eq:D-B}
D=\frac{1}{2\gamma }\xi^2(x) B \xi^{-2}(x)
\eeq
where the function $\xi (x)$ is given by:
\beq
\label{eq:xiJ}
\xi (x)=\prod_{i<j}|\sinh^2 x_j-\sinh^2
x_i|^{\frac{\beta}{2}}\prod_i|\sinh 2x_i|^{\frac{1}{2}}~. \eeq
and the operator $B$ depends only on the radial coordinates $\{x_i\}$
of the space according to the relation:
\beq B=[\xi(x)]^{-2}\sum_{i=1}^N\frac{\de}{\de x_i}[\xi(x)]^{2}
\frac{\de}{\de x_i}\eeq

As a consequence, if $\Phi_k(x)$, $x=\{x_1,\cdots,x_N\}$,
$k=\{k_1,\cdots,k_N\}$ is an eigenfunction of
$B$ with eigenvalue $k^2$, then $\xi(x)^2\Phi_k(x)$ will be an
eigenfunction of the DMPK operator with eigenvalue $k^2/(2\gamma)$.

Once the eigenfunctions of the DMPK operators are known it is rather
straightforward to construct the Green function and use it to solve the 
DMPK equation. The solution turns out to be rather involved, but it drastically
simplifies in the two interesting limits of insulating (small values of $k$) and
metallic (large values of $k$) regimes. It is important to stress at this point
that even if an explicit form is known only for the special case $\beta=2$, 
a general and powerful
 asymptotic expansion for large values of $x$ exists for all the possible
 symmetric spaces~\cite{caselle1,review}. This asymptotic expansion 
 turns out to be
 enough for the construction of the two (insulating and conducting) limiting 
 solutions. Its explicit form is:

\begin{equation}
\Phi_k(x)\sim \frac{1}{\xi(x)} \left(\sum_{r\in W} c(r k) e^{i(r k,x)}
\right)~~,
\label{ss3}
\end{equation}
where $r k$ is the vector obtained acting with $r\in W$ on $k$,
 $W$ is the Weyl group associated to the root system of the
symmetric  space and $(k,x)$ denotes the scalar product.

All the informations related to the underlying
symmetric space are encoded in the function $c(k)$ which is given by:
\eq
c(k)=\prod_{\alpha\in\Delta^+} c_\alpha(k)
\en
with
\beq c_\al(k)=\frac{\Gamma (i(k,\al)/2)}{\Gamma (m_\alpha /2
+i(k,\al)/2)}\eeq 
where
$\Gamma$ denotes the Euler gamma function,
$\alpha$ is a generic root belonging to the root lattice which defines the
symmetric space, $m_\alpha$ denotes its multiplicity 
and the product is
restricted to the sublattice $\Delta^+$ of positive roots only.

\section{"Clustered" solutions of the DMPK equation}
In the previous section the integrability of the DMPK equation has been stressed
in order to write explicitly, at least in the asymptotics, its solutions.
In several physical applications (see next section) it would be interesting to
study  solutions of the DMPK equation in which the symmetry among the $N$
eigenvalues is broken.
The aim of this paper is to show how to exploit again the integrable nature of the
DMPK equation in order to go beyond the isotropic solutions obtained in \cite{caselle1}.
The key point is that, given the integrability of the DMPK equation, its solution can
 be written even if non trivial initial conditions are assumed; this observation allows to construct non isotropic solutions starting from an equation
 which is isotropic by construction. 
 The simplest way to obtain this is to impose that some of
the eigenvalues form a ``cluster'' (i.e. the distances among them remain finite
while the distances with respect to the other ones go to infinity), or more
generally a set of independent clusters; the peculiar form of the DMPK equation
 ensures that such  solutions exist and
survive in the asymptotic limit. These solution obviously require 
 suitably chosen initial conditions, which are degenerate in the whole phase
 space. 
The main goal of this paper is to show that also these clustered
 solutions admit
 an asymptotic expansion similar to that of eq.(\ref{ss3}). 
Given this
 expansion,  one can then obtain the probability density
 $P(\{x_n\},s)$ in presence of these clusters
in a way analogous to the isotropic case.
 We shall consider below some possible
 applications of this result.

 Let us see these solutions in more detail.
Let us assume the cluster to be composed by the first $N'<N$ eigenvalues. 
This means:
\bea  |x_i-x_{j}|< \infty, &i,j=1,\ldots , N'~~~~(i\not=j).
\\\nonumber  |x_i-x_{j}|\rw \infty,& i=1,\ldots ,
N;~~~j=N'+1,N~~ (i\not=j) \label{condcluster}\eea
In the symmetric space framework we can identify the cluster by selecting
a subset of
the root system associated to the space. Let $\Pi$ be the system of
simple roots associated to $\x$, and $\Pi'$ be a subsystem of simple
roots which satisfies the inequality:
\beq\label{clustercond} 
\Pi'=\{\al \in \Pi /\lim_{|x|\rw \infty} x_\al <\infty\}
\eeq
where $x_\al=(x,\al)$.

There are at this point two possibilities. Since $\Pi$ is a $C_N$ type lattice 
then  $\Pi'$ can be again of type $C_N$ (in this case it must also contain 
the long root and the ordinary roots must be chosen so as to keep the $Z_2$
symmetry of the lattice) or it can
be of type $A_N$. In both cases from the ordinary roots of
 $\Pi'$ one can construct the
differences $x_\al=x_i-x_{i+1}$ which correspond to the nearest
neighbours distances between some of the eigenvalues. From the definition of
$\Pi'$ these
distances must remain finite in the asymptotic limit  so that
 $\Pi'$ defines a cluster (if it is connected)
or a set of clusters otherwise. If the cluster is of type $A_N$ there is no
other costraint and the cluster can in principle flow to an infinite distance
from the origin (while keeping a finite distance among the eigenvalues inside
the cluster). If the
cluster is of type $C_N$ on the contrary the eigenvalues are 
 bounded (by the lattice structure
 of $\Pi'$ itself) to stay within a finite distance from the origin.
 In the following we shall denote with
$\ti x$  the set of radial coordinates outside the cluster and with
$x'$ the ones inside the cluster.

The asymptotic expansion of the ZSF's in presence of such  a cluster 
was obtained a few years ago by Olshanetsky in \cite{olshanetsky2}. It turns out
to be a rather natural generalization of the Harish-Chandra result of 
eq.(\ref{ss3}):
\beq
 \Psi_k(x)\sim \sum_{r\in W/W'}
c_z(\tilde{rk})e^{i(\tilde{rk},\ti x)}\Psi_{(rk)'}(x')\label{eq:cluster}
\eeq
where $\Psi_k(x)$ is given by the product $\xi(x)\Phi_k(x)$. In this formula 
$(rk)'$ denotes the projection of the vector $rk$ on the sublattice
$\Pi'$ and $\tilde{rk}$ its complement, $\xi(x)$ is given by eq.(\ref{eq:xiJ})
and 
\beq c_z(k)=\prod_{\al\in
\Delta^+/\Delta^{'+}}c_\al(k)\eeq
where $\Delta^{'+}$ is the set of positive roots associated to the cluster, and
the $W'$ which appears in the coset
$W/W'$ is the Weyl group associated to the cluster. This expression is
apparently simple but it is highly non trivial.
Notice for instance that the symmetrization with respect to  
the Weyl coset $W/W'$ acts not only on the part containing
the coordinates $\ti x$ but also on the momenta of the zonal spherical function
describing the cluster coordinates 
$\Phi_{(rk)'}(x')$.
This means that the particles inside the cluster do not move independently in 
a section of the whole space but they ``feel'' the presence of the other particles
and are subject to the symmetry group of the remaining space.
It is interesting  to notice that the above construction could be formulated
also in the framework of the original Transfer Matrix ensemble (i.e. before
diagonalizing the transfer matrix). The clustered solutions correspond in this
framework to peculiar {\bf boundary} limits of the original symmetric spaces
(which are known in the mathematical literature as Martin
boundaries~\cite{martin,ol3}).
The precise characterization of these boundaries is outside the scope of the
present paper. We plan to address this issue in a future publication.

Let us see two examples which may hopefully clarify the issue:

\vskip 0.1cm
{\bf Example 1}: a $C_N$ type cluster composed by the first
 $(N-1)$ eigenvalues plus an
isolated eigenvalue which flows to infinity, i.e. (see eq.(\ref{condcluster}))
\bea  |x_i-x_{j}|< \infty, &i,j=1,\ldots , N-1~~~~(i\not=j).
\\\nonumber  |x_i-x_{N}|\rw \infty,& i=1,\ldots ,
N-1 \nonumber\eea
Then the asymptotic form of the ZSF, 
and  of the related product function $\Psi_k(x)$ is:
\eq
\Psi_k(x)\sim\sum_{j=1}^N
c_{z}(k_j)\Psi_{\hat{k_j}}(x')e^{ik_j x_N}
\en
where $\Psi_{\hat{k_j}}(x')$ is associated to the ZSF of the 
$C_{N-1}$ symmetric space of the
cluster, $\hat{k_j}\equiv (k_1,\cdots k_{j-1},k_{j+1},\cdots k_N)$ 
denotes the collection of $(N-1)$ momenta in which $k_j$ is omitted and
$c_z(k_j)$ is
given in this case by the product of the $c_\alpha(k)$ functions over all the
(positive) roots of type $\alpha=\pm e_j \pm e_l~~(\forall l\not=j)$,  
$\alpha=2e_j$~~($j$ fixed).
 From this
expression it is straightforward to construct iteratively the ZSF in which two or
more eigenvalues flow to infinity.
\vskip 0.1cm
{\bf Example 2}: a $C_N$ type cluster composed by the first
 two eigenvalues: $x_1$ and $x_2$, while the other $(N-2)$ flow to infinity. 
\eq
\Psi_k(x)\sim\sum_{s\in{\hat W}}c_z(k)\Psi_{{s(1),s(2)}}(x_1,x_2)e^{i\sum_{r=3}^N k_{s(r)} x_r}
\en
where $\hat W\equiv W/W'$ 
is the coset of Weyl groups associated to the whole set of eigenvalues and to
the cluster respectively, the subscript of the two-particle wavefunction
$\Psi_{{s(1),s(2)}}(x_1,x_2)$ identifies the two momenta to which this function
 is associated and
$c_z(k)$ in this case is the product of the $c_\alpha(k)$'s over all the
positive roots of
type $2e_{s(m)},~~2e_{s(n)}$ (with $m,n>2$) 
and $\pm e_{s(m)}\pm e_{s(n)}$ (with the only exclusion of the combination
$(m,n)=(1,2)$ or $(2,1)$~).

\section{Applications}
Among the various possible uses of these new solutions we see in particular
four interesting applications. 

\vskip 0.1cm
{\bf 1]} We can use them to model  systems in which the number of open channels
 is reduced by the structure of the wire itself (see the wide-narrow-wide geometry 
of \cite{beenakker3})\footnote{We thank C.W.Beenakker for suggesting us this
possibility.}.
 In this case, one could consider a
configuration formed by a $C_N$ type cluster (bounded to the origin) made of
$N'$ eigenvalues  and let the remaining $N-N'$ eigenvalues flow to infinity (see
example 1 above).
One can choose $N'$ and $s$ (the reduced length of the wire) so as to keep the cluster
in the metallic regime, while the other eigenvalues are in the insulating one
and do not contribute to the wire conductance. 

{\bf 2]}
 The same configuration discussed above could be used to provide a simple 
 but effective way to improve our description of real quantum wires.  
While in the previous example $N'$ was fixed (being the number of the open
channels in
the narrow part of the wire) one could easily generalize the example 
keeping $N'$ as an additional
  degree of freedom  which can be varied so as 
  to take into account the effect of external parameters like the amount of
  disorder in the wire.
The rationale behind this
proposal is the well known idea~\cite{imry} that in a generic conductor only a
fraction (which depends on the disorder) of its channels is open and the
isotropy anzatz should be a very good approximation for them, while the
non-trivial (i.e. non one dimensional) behaviour of the conductance is due to
the variation in the number of open channels.

{\bf 3]} The previous  picture could be refined choosing configurations with more than one
cluster, each cluster being characterized by a different distance at which the
crossover between metallic and insulating regime occurs (fixed by the only free
parameter which we have in these generalized solutions, that is the number of
eigenvalues of each cluster).
This multi-cluster structure could be used in order to evade the isotropy ansatz (this was indeed the main
reason for the present investigation).
In order to obtain  the multi-cluster configuration it is sufficient to choose 
a {\em not connected} subset of roots $\Pi'$ which satisfies the condition 
given in eq. (\ref{clustercond}). 
The roots belonging to  $\Pi'$ correspond to a set 
of coordinates $x'$  which are grouped into  different clusters;
the number of clusters $N_c$ and their width $N'_i$ are uniquely  determined by
the choice of the not connected root lattice $\Pi'$.
The asymptotic expansion given in the general form of eq. (\ref{eq:cluster})
 can then be applied in order to find the proper solutions of this 
multi-cluster configuration.

The isotropic symmetric space characterized by $N$ radial coordinates in this case is mapped 
into an anisotropic model in which the $N_c$ channels correspond to  $N_c$ clusters (each cluster 
containing a different number $N_i'$  of coordinates); this reconfiguration should ensure a
 different probability for the electron to be scattered among the different channels.
It is important to stress once again that in our proposal this goal is reached by properly choosing 
the initial
conditions, while the DMPK equation (with all
its remarkable properties) {\sl is kept unchanged}. This means, as a side remark, that the DMPK equation
 still depends on only one parameter i.e. 
the ratio $N/s$. The $N_c$  parameters which appear in the special solutions in which we are interested 
(see the general solution and the examples discussed in sect.4) come from the initial conditions and simply
consist in the list of the sizes $N_i'$ of the $N_c$ clusters. 
In this type of solutions the $N_c$ channels  will encounter 
the metal-insulator transition at different lengths of the wire; this feature
 suggests that the $N_i'/s$ parameters can be related to the different mean free paths associated to the 
 channels. This picture reminds  the ``non equivalent channels model'' proposed by Mello and Tomsovic
  in \cite{mello} but differs from it since our proposal (as mentioned above) does not require to modify
    the DMPK equation. 

\vskip 0.1cm
{\bf 4]} An interesting independent application is related to the recent attempt
~\cite{gm,gmw} to describe the crossover between the metallic and insulating regime
in quasi 1D systems by separating out the first~\cite{gm} or the first
two~\cite{gmw} eigenvalues and considering the rest as a continuum in the solution
of the DMPK equation. This approximation reminds the clustered solutions that we
have discussed in this paper (see in particular example 2). The fact that these
clustered configurations are indeed {\sl exact} solutions of the DMPK equation
(albeit with non trivial initial conditions) may explain the remarkable
stability of the saddle point solution noticed in~\cite{gmw} and could offer an
independent justification for the approach.

\vskip 1cm
{\bf  Acknowledgements}
This work was partially supported by the 
European Commission TMR programme programme HPRN-CT-2002-00325 (EUCLID).


\begin{thebibliography}{9}

\bibitem{dmpk} Dorokhov O. N., Pis'ma Zh. Eksp. Teor. Fiz. 36
(1982) 259; Mello P. A., Pereyra P., Kumar N., Ann. Phys. 181
(1988) 290

\bibitem {caselle1} 
M.~Caselle,
Phys.\ Rev.\ Lett.\  {\bf 74} (1995) 2776.
M.~Caselle, 
Nucl.Phys.Proc.Suppl. 45A (1996)120-129
\bibitem{review} For a review see for instance:
 M. Caselle and U. Magnea
{\ttfamily cond-mat/0304363}
\bibitem{br} C.~W.~J.~Beenakker and B.~Rejaei, Phys. Rev. Lett.
{\bf 71}, 3689 (1993); Phys. Rev. B {\bf 49}, 7499 (1994).


\bibitem{beenakker1} Beenakker C.W.J.
    Rev. Mod. Phys. 69 (1997) 731

\bibitem{mwgg}   Muttalib K. A.,  Woelfle P.,  Garcia-Martin A., and  Gopar V. A., Europhys. Lett. {\bf 61}, 95 (2003) )



\bibitem{zzz}
A. Garcia-Martin, J.J. Saenz, Phys. Rev. Lett. 87, 116603 (2001);
P. Markos, Phys. Rev. B {\bf 65} 104207 (2002);
 M.R\"uhl\"ander and C. Soukoulis, Physica B Cond. Matt. {\bf 296},32 (2001);
 M. R\"uhl\"ander, P. Markos and C. M. Soukoulis, Phys.Rev. B {\bf 64} 
 212202 (2001).
\bibitem{gm} 
Muttalib K. A. and W\"olfle P., \PRL{83} (1999) 3013; Ann. Phys. (Leipzig)
 {\bf 8} (1999) 753. 
\bibitem{gmw}
Muttalib K. A., Gopar V.A. and W\"olfle P., Phys. Rev {\bf B66}
(2002) 174204.
\bibitem{mello}  Mello P.A., Tomsovic S., Phys. Rev. Lett. {\bf 67}, 342 (1991),
 Mello P.A., Tomsovic S., Phys. Rev. B {\bf 46}, 15963 (1992))

\bibitem{yyy} K.A. Muttalib and V.A. Gopar Phys. Rev. B 66 115318 (2002);
K. A. Muttalib and J. R. Klauder, Phys. Rev. Lett. {\bf 82},4272 (1999);
P. Markos, Phys. Rev B {\bf 65}, 092202 (2002);
J. T. Chalker and M. Bernhardt, Phys. Rev. Lett.{\bf 70}, 982 (1993).

\bibitem{bk} F. A. Berezin and F. I. Karpelevick, Dokl. Akad. Nauk SSSR
{\bf 118}, 9 (1958).
\bibitem{hc} Harish-Chandra, Amer. J. Math {\bf 80} 241 (1958);
{\bf 80} 553 (1958).

\bibitem{olshanetsky} Olshanetsky M. A., Perelomov A. M., Phys. Rep. 94
(1983) 313
\bibitem{landauer} Landauer R., Philos. Mag. {\bf 21} 863 (1970).

\bibitem{olshanetsky2} Olshanetsky M. A., \TMP{95}  (1993) 341
\bibitem{martin} Martin, R.S., Trans. Amer. Math. Soc. {\bf 49} (1941) 137.
\bibitem{ol3} Olshanetsky M. A., Usp. Math. Nauk. {\bf 24}  (1969) 189
\bibitem{beenakker3} Beenakker C.W.J., Melsen J.A., Phys.Rev.B {\bf 50} 2450 (1994)
\bibitem{imry} Y. Imry, Europhys. Lett. {\bf 1} 249 (1986).
\end{thebibliography}
\end{document}